\pdfoutput=1

\documentclass[twocolumn,aps,prb,superscriptaddress,longbibliography]{revtex4}
\usepackage{amssymb,amsfonts,amsmath,amsthm}
\usepackage{color,graphicx,hyperref,mathrsfs}
\usepackage{enumerate}
\usepackage[normalem]{ulem}
\usepackage{siunitx,upgreek}

\newcommand{\R}{\mathbb R}

\newcommand{\iks}{s}
\newcommand{\m}{\mathbf{m}}
\newcommand{\mpa}{m_\parallel}

\newcommand{\mpe}{\mathbf{m}_\perp}

\begin{document}
\title{Asymptotically exact formulas for the stripe domains period in
  ultrathin ferromagnetic films with out-of-plane anisotropy}

\author{Anne Bernand-Mantel}
\email{anne.bernand-mantel@cemes.fr}

\affiliation{CEMES, Université de Toulouse, CNRS, 29 Rue Jeanne
  Marvig, BP94347, 31055 Toulouse, France}

\author{Valeriy V. Slastikov}

\affiliation{School of Mathematics, University of Bristol, Bristol BS8
  1UG, United Kingdom}

\author{Cyrill B. Muratov}

\affiliation{Dipartimento di Matematica, Universit\`a di Pisa, Largo
  B. Pontecorvo, 5, 56127 Pisa, Italy}

\date{\today}


\begin{abstract}
  We derive asymptotically exact formulas for the equilibrium magnetic
  stripe period in ultrathin films with out-of-plane anisotropy that
  include the full domain wall magnetic dipolar energy. Starting with
  the reduced two-dimensional micromagnetic model valid for thin
  films, we obtain the leading order approximation for the energy per
  unit volume in the vanishing film thickness limit in the case of
  Bloch and Néel wall rotations. Its minimization in the stripe period
  leads to an analytical expression for the equilibrium period with a
  prefactor proportional to the Bloch wall width. The constant in the
  prefactor, related to the long-range dipolar interactions, is
  carefully evaluated. This results in a remarkable agreement of the
  stripe domain energy density and stripe period predicted by our
  analytical formulas with micromagnetic simulations.  Our formula can
  be used to accurately deduce magnetic parameters from the
  experimental measurements of the stripe period and to systematically
  predict the equilibrium stripe periods in ultrathin films.
\end{abstract}
\maketitle
\section{Introduction}
\label{sec:introduction}


Magnetic domain observation and prediction  have been the subject of
continuous scientific interest for over a hundred
years\cite{hubert}. The characteristics of magnetization patterns are
notoriously difficult to elucidate, since they intricately depend on
the balance of multiple interactions present in magnetic systems. A
particular challenge comes from the fact that magnetic systems fall
into the category of systems with competing short-range and long-range
interactions\cite{hubert,landau8,m:pre02}. Specifically, magnetic
stripes and bubbles in ferromagnetic films with out-of-plane
anisotropy are a prime example of the magnetic domain patterns forming
as a result of such a competition. They were first observed at the end
of the 50's and were the subject of extensive studies related to their
application in magnetic bubble memories\cite{bobeck72,malozemov79}. In
the late 80's, the progress in deposition techniques enabling
atomically resolved growth of ultrathin film multilayers led to a
revival in magnetic domain observations in thin films with
out-of-plane anisotropy and to the discovery of properties specific to
the nanoscale. These lower-dimensional magnetic systems gave rise to
multiple new opportunities in both fundamental and applied
magnetism\cite{himpsel98}.

In magnetic thin films the magnetostatic interaction tends to maintain
the magnetization in the film plane \cite{aharoni}. However, it was
observed that for films of thicknesses of the order of a few
nanometers or less, the anisotropy of interfacial origin may promote
an out-of-plane magnetization in transition metal
systems\cite{gradmann68}. Magnetic stripe patterns were observed in
multilayers\cite{draaisma87,barnes94,johnson96} and ultrathin films
with thicknesses tuned close to the spin reorientation
transition\cite{allenspach92,vaterlaus00,wu04}. More recently, the
presence of a Dzyaloshinskii-Moriya interaction (DMI) of interfacial
origin in ultrathin films was brought to light
\cite{crepieux98,bode07}. Chiral N\'eel stripes \cite{chen13} and
skyrmionic bubbles\cite{jiang15} were observed in ultrathin films. The
possibility to manipulate skyrmions by currents \cite{jiang15} and
electric fields \cite{schott17}, and the associated potential of
skyrmions for information technology applications
\cite{tomasello14,prychynenko18} led to a second revival in
experimental and theoretical investigations on magnetic stripes in
ultrathin magnetic films and multilayers.

Modeling of magnetization patterns in thin films requires to evaluate
precisely the demagnetizing energy, a notoriously difficult task owing
to the slow convergence and often singular behavior of the
demagnetizing energy integrals in space due to the long-range nature
of the magnetostatic interaction. For example, it is not possible to
calculate exactly the demagnetizing energy of an isolated 1D domain
wall in a film of infinite extent in the plane, as the respective
integrals diverge\cite{desimone06}. In the ultrathin film limit, the
magnetic surface and volume charges decouple\cite{kmn:arma19} and the
full demagnetizing energy including long-range dipolar interaction can
be calculated explicitly in the case of a compact magnetic skyrmion,
using Fourier transform\cite{bernand-mantel20,bernand-mantel23}. In
the case of stripe domains, the periodic character of the pattern
enables the use of Fourier series. This feature resulted in various
modeling studies of stripe patterns in a countless number of
publications, among which only a few will be cited here.

Theoretical studies of stripes started with the work of Kittel, who
treated the case of equilibrium stripe periods much smaller than the
film thickness\cite{kittel46}. This was followed by the work of Maleck
and Kambersk\'y \cite{malek58}, who studied the case of thicknesses of
the order or lower than the domain width and predicted an increase of
the stripe period with the decrease of the film thickness under a
certain threshold thickness. Kooy and Enz proposed a generalized
demagnetizing energy applicable to a wide range of
thicknesses\cite{kooy60}.

In the case of a magnetic monolayer, Yafet and Gyorgy included
explicitly the domain wall dipolar energy and predicted a striped
ground state for sufficiently large anisotropy\cite{yafet88}. Czech
and Villain considered a regime in which a monolayer may exhibit an
exponential dependence of the domain size on the inverse strength of
the dipolar interaction\cite{czech89}, a behavior that was already
identified for the stripe patterns in the context of Langmuir
monolayers, whose modeling bears many similarities to magnetic
systems\cite{andelman85}.  Kaplan and Gehring\cite{kaplan93} proposed
a micromagnetic model in which the domain wall thickness is neglected
and obtained a semi-analytical formula predicting a stripe period with
an exponential dependence on the inverse film thickness. This formula
was later derived explicitly by Millev\cite{millev96}, who obtained an
analytical expression for the prefactor under the same assumptions. A
similar semi-analytical formula for a two-dimensional Ising
ferromagnet on a square lattice with dipolar interactions was obtained
by MacIsaac et al. \cite{macisaac95}. In parallel, Kashuba and
Pokrovsky \cite{kashuba93}, and later Sukstanskii and Primak
\cite{sukstanskii97}, took into account the finite domain wall width
in an ultrathin film and derived an analytical formula, where the film
thickness present in the Kaplan and Gehring formula in the prefactor
is replaced by the Bloch wall width.

More recently, studies of the stripe problem took into account the
Néel character of the domain wall that results in an additional
nonlocal dipolar energy term related to the volume magnetic
charges. Lemesh et al.  \cite{lemesh17} derived an expression for the
energy whose numerical minimization gives a prediction of the stripe
period, as well as the domain wall width and angle. Meier et
al. proposed a modified version of the Kashuba and Pokrovsky formula
taking into account the Néel character of the domain
wall\cite{meier17}.

Despite the extensive efforts in modeling the stripe patterns in
magnetic thin films, at present there exists no clarity about the
regime of validity of various existing analytical formulas for the
equilibrium stripe periods. Formulas with a prefactor proportional to
the film thickness
\cite{hoffmann02,bergeard10,schott17,je18,bouard18,schott21,balan23}
or the Bloch wall width \cite{ando16,dohi16,meier17,yang21} are used
indiscriminately for ultrathin films, despite presenting up to more
than one order of magnitude difference\cite{skomski98}. Additionally,
even among the models which take into account the domain wall
long-range dipolar contributions in the stripe energy, resulting in an
analytical formula with a prefactor proportional to the Bloch wall
width\cite{kashuba93,sukstanskii97, skomski98,meier17}, there exist
discrepancies in the estimated constants in front of the prefactor up
to a factor of 3. This is problematic, as these formulas often serve
as a tool to quantitatively estimate the domain wall energy from the
stripe period.

In the present work we settle this issue by calculating an
asymptotically exact analytical formula for the equilibrium magnetic
stripe period valid in the ultrathin film regime, were the prefactor
is calculated explicitly for the case of both Bloch and Néel domain
walls. The validity of the obtained formulas is tested using detailed
micromagnetic simulations, in which we compare the energy density
scaling with the domain period, as well as the domain period variation
with the film thickness for systems with and without DMI. Our results
confirm that for ultrathin films the full domain wall contribution to
the energy needs to be taken into account, leading to a
proportionality of the prefactor to the Bloch wall width and providing
a prefactor that is asymptotically exact for vanishing thicknesses.

Our paper is organized as follows. In Sec. \ref{sec:summary}, we
present the formulas for the energy per unit volume and the
equilibrium period of magnetic stripe domains derived by us, in a
dimensional form convenient to use for comparison with experiments. In
this section, we also compare the predictions of our formulas with the
results of direct numerical simulations for several sets of
experimentally relevant parameters and compare our findings with
previous literature. The rest of the paper provides the details of our
derivation and the simulations. Specifically, in Sec. \ref{sec:Model},
we set up a reduced mathematical model whose solution asymptotically
governs the energetics of the equilibrium stripe domains. In
Sec. \ref{sec:calculation}, we carry out an asymptotic analysis of
this model to extract the leading order expansion for the energy
density and the equilibrium stripe period. Finally, in
Sec. \ref{sec:micromag} we present the details of our numerical
simulations and in Sec.\ref{sec:concl} we draw conclusions.

\section{Summary of the results}
\label{sec:summary}

\begin{figure}[t]
  \centering
\includegraphics[width=8cm]{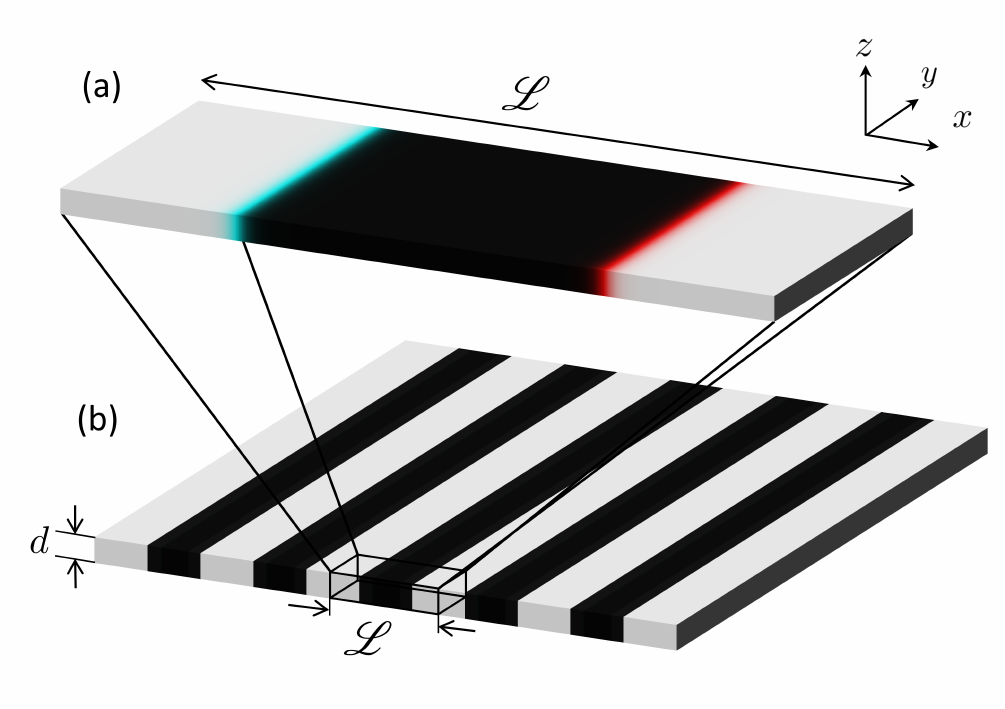}
\caption{Schematic representation of an ultrathin film of thickness
  $d$ with periodic stripe magnetic domains of period ${\mathscr
    L}$. The black and white colors represent magnetization pointing
  up and down in the $z$-direction, while blue and red represent the
  in-plane magnetization pointing left and right in the $x$-direction
  (Néel domain walls).}
  \label{fig:Schem}
\end{figure}

We consider a sample of infinite extent in the plane that consists of
a thin ferromagnetic film with thickness $d$ and out-of-plane magnetic
anisotropy defined by a volume uniaxial anisotropy constant $K_u$. We
define the dimensionless film thickness
$\delta = d / \ell_\mathrm{ex}$, where
$\ell_\mathrm{ex} = \sqrt{2 A_\mathrm{ex} / (\mu_0 M_s^2)}$ is the
exchange length, $A_\mathrm{ex}$ the exchange stiffness, $\mu_0$ the
vacuum permeability and $M_s$ the saturation magnetization. The film
is assumed to be sufficiently thin, $\delta \lesssim 1$, in order for
the magnetization vector $\mathbf M$ to be independent of the
$z$-variable (see Fig.~\ref{fig:Schem}). We introduce the
dimensionless DMI strength $\kappa = D / \sqrt{A_\mathrm{ex}K_d}$,
where $K_d= \frac12 \mu_0 M_\mathrm{s}^2$ and $D$ is the interfacial
DMI constant (in J/m$^2$, non-negative without loss of generality). It
is widely expected that for $\delta$ and $\kappa$ sufficiently small
the magnetic ground state of the sample consists of periodic parallel
magnetic stripes with a period ${\mathscr L}$ (see
Fig.~\ref{fig:Schem}), where the stripes are magnetic domains of width
$\simeq \frac12 \mathscr L$ with magnetization pointing alternatively
up and down depending only on $x$, separated by domain walls. We
assume that the magnetization vector rotation lies in a plane: either
the $xz$-plane for N\'eel stripes or the $yz$-plane for Bloch stripes.
  
To illustrate our results, we consider the case of a thin transition
metal ferromagnetic film with an exchange constant $A_\mathrm{ex}=10$
pJ/m and saturation magnetization $M_s=1$ MA/m, resulting in an
exchange length $\ell_\mathrm{ex} \simeq 4$ nm.

\subsection{Bloch stripes}
\label{subsec:Bloch}

In Sec. \ref{sec:bloch_calc}, we compute the energy per unit volume
${\mathcal F}({\mathscr L}) = f(\mathscr L / \ell_\mathrm{ex}) K_d$,
after subtracting the energy of the uniformly magnetized state, of
periodic stripe domains of period $\mathscr L$ in the classical case
of stripes separated by Bloch walls ($D=0$). We obtain
\begin{align}
  {\mathcal F}({\mathscr L}) \simeq  {2 \sigma_{B}\over
  {\mathscr L}}  - {4 d K_d \over \pi {\mathscr L}} 
  \left[ \ln\left(\frac{{\mathscr L}}{\pi^2L_B}\right) + \gamma
  + 1
  \right],
\end{align}
where $\sigma_{B}=4\sqrt{A_\mathrm{ex}K_\mathrm{eff}}$ is the wall
surface tension, in which $K_\mathrm{eff}=K_u-K_d > 0$ is the
effective magnetic anisotropy of the thin film forcing the
magnetization to align normally to the film plane,  
$L_B= \sqrt{A_\mathrm{ex} / K_\mathrm{eff}}$ is the Bloch wall width,
and $\gamma \approx 0.5772$ is the Euler-Mascheroni constant.  We
obtain the leading order optimal period by minimizing
${\mathcal F}({\mathscr L})$ in ${\mathscr L}$:
\begin{align}
  \label{LoptB}
  {\mathscr L}_\mathrm{opt}^\mathrm{Bloch} = \pi^2
  e^{-\gamma} L_B \exp \left( { \pi \sigma_{B} \over\mu_0 M_s^2d}
  \right).
\end{align}
This formula is asymptotically exact for $d \to 0$. The formula is,
therefore, expected to give a good approximation to the optimal period
as soon as $d \lesssim L_B$ and $\sigma_B \gtrsim d K_d$.
 
\begin{figure}[t]
\includegraphics[width=8.75cm]{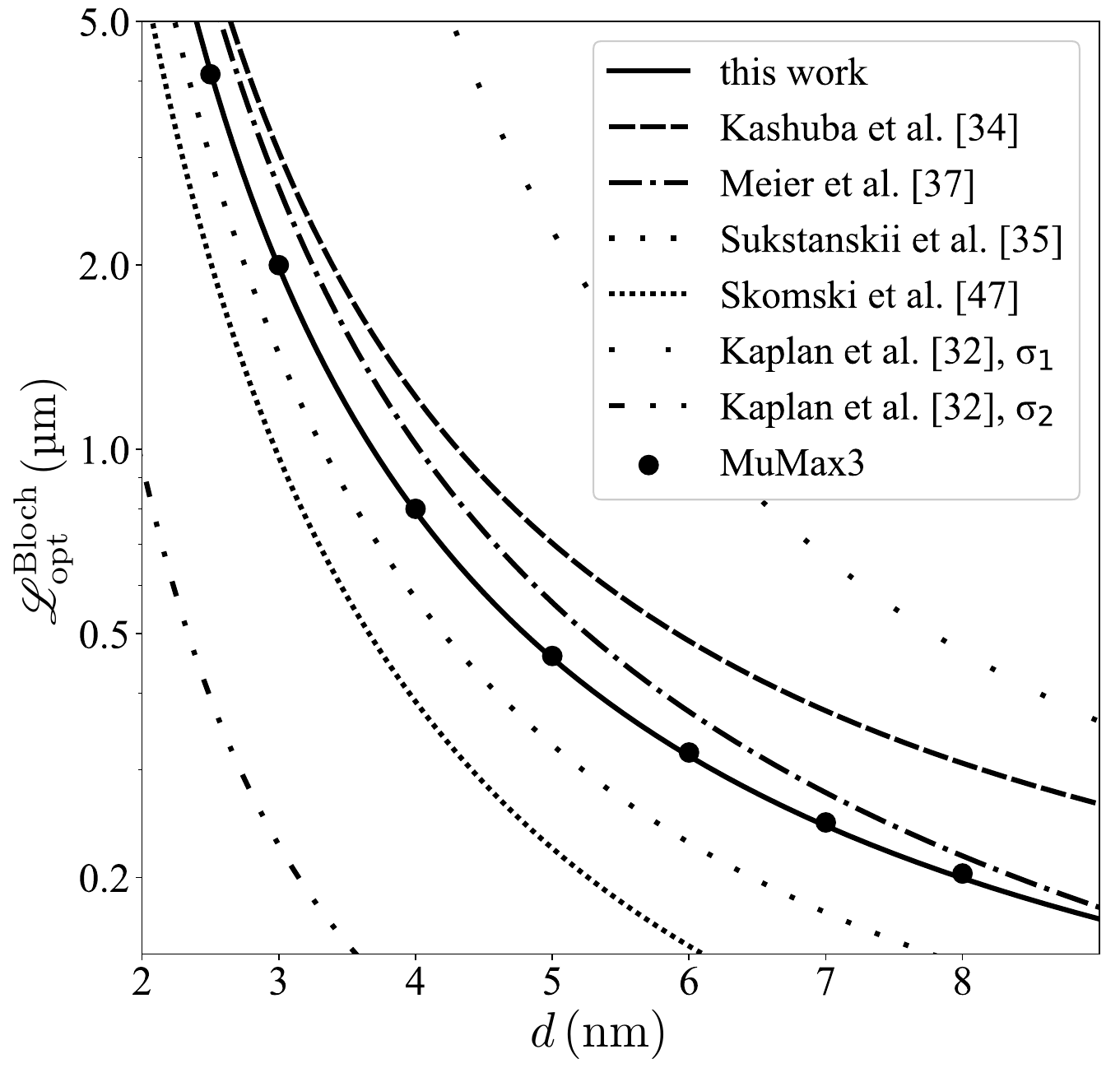}
\caption{Equilibrium stripe period
  ${\mathscr L}_\mathrm{opt}^\mathrm{Bloch}$ as a function of the film
  thickness $d$ in the regime of Bloch walls (zero DMI). The material
  parameters are $A_\mathrm{ex}=10$ pJ$/$m, $M_s=1$ MA$/$m and
  $K_{u1}=0.75$ MJ$/$m$^3$ (regime 1). The asymptotic equilibrium
  stripe period in Eq. \eqref{LoptB} is represented by a solid
  line. The equilibrium stripe period obtained by micromagnetic
  simulations using {\sc MuMax3}\cite{vansteenkiste14} (see
  Sec.~\ref{sec:micromag} for details) is represented by black dots.
  The equilibrium stripe periods from previous
  works\cite{kashuba93,sukstanskii97,skomski98,meier17, kaplan93} are
  also shown (see inset for details).}
\label{fig:Bloch}
\end{figure}

In order to illustrate the Bloch stripes period formula, we chose a
volume magnetocrystalline anisotropy tuned to $K_{u1}=0.75$ MJ/m$^3$
(regime 1) and independent of the film thickness (the regime of
magnetocrystalline anisotropy of surface origin will be considered in
Sec.~\ref{subsec:Neel}).  In Fig.~\ref{fig:Bloch}, we present the
optimal stripe period for regime 1, where the prediction from
Eq. \eqref{LoptB} is represented as the solid line. We use black dots
to present the optimal period obtained using micromagnetic simulations
using the {\sc MuMax3} software\cite{vansteenkiste14}. These numerical
results, obtained with the help of the procedure described in detail
in Sec.~\ref{sec:micromag}, were computed by minimizing the energy of
one stripe with period $ {\mathscr L}$, under periodic boundary
conditions (see Fig.~\ref{fig:Schem}). The stripe period is varied
until the optimal stripe period, corresponding to the smallest
minimized energy density, is found. The micromagnetic simulations are
in a very good agreement with our asymptotic formula in
Eq.~\eqref{LoptB} for the values of $\delta$ up to $\delta=2$ (see
Fig.~\ref{fig:Bloch}). The variation of the equilibrium Bloch stripe
period as a function of magnetocrystalline anisotropy for a fixed
thickness can be found in the supplemental material \cite{suppl}.

\subsection{Néel stripes}
\label{subsec:Neel}
We now consider a system with DMI, where
${4 \ln 2 \over \pi^2} d K_d < D < {4 \over \pi} \sqrt{A _\mathrm{ex}
  K_\mathrm{eff}}$, i.e., for $D > 0$ sufficiently large to ensure a
pure Néel rotation for the wall, but sufficiently small to prevent
spin spirals\cite{thiaville12,lemesh17}.  In Sec. \ref{sec:neel_calc},
we compute the energy per unit volume
${\mathcal F}({\mathscr L}) = f(\mathscr L / \ell_\mathrm{ex}) K_d$,
after subtracting the energy of the uniformly magnetized state, of
periodic stripe domains of period $\mathscr L$ in the case of Néel
walls. We obtain
\begin{align}
  {\mathcal F}({\mathscr L}) \simeq  {2 \sigma_{N}\over
  {\mathscr L}}  - {4 d
  K_d \over \pi {\mathscr L}} \left[\ln\left(\frac{{\mathscr
  L}}{2\pi^2L_B}\right) +\gamma+1 \right],
\end{align}
where $\sigma_{N}=4\sqrt{A_\mathrm{ex}K_\mathrm{eff}}-\pi D > 0$ is
the surface tension of the N\'eel domain wall. Minimizing this energy
in ${\mathscr L}$, we obtain
\begin{align}
  \label{LoptN}
  {\mathscr L}_\mathrm{opt}^\mathrm{N\acute{e}el}  =2 \pi^2
  e^{-\gamma} L_B \exp  
  \left( { \pi \sigma_{N} \over\mu_0 M_s^2d}  \right).
\end{align}
Again, this formula is asymptotically exact for $d \to 0$ and is,
  therefore, expected to give a good approximation to the optimal
  period as soon as $d \lesssim L_B$ and $\sigma_N \gtrsim d K_d$.

This expression differs in two points from that in
Eq. \eqref{LoptB}. First, the expression for the domain wall surface
tension appearing in the exponential factor is, as expected, taking
into account a decrease in the surface tension by $\pi D$ due to
DMI\cite{thiaville12,ms:prsa17}. Second, a factor of two, related to
the nonlocal dipolar interactions is appearing in the expression for
the prefactor in the case of Néel walls. This additional energy is a
consequence of the volume charges present in the Néel wall, associated
with the in-plane component of the demagnetizing field, whose
analytical expression in the limit $d \to 0$ is a well-known result
\cite{tarasenko98,thiaville12,meier17}.

\begin{figure}[t]
\includegraphics[width=8.5cm]{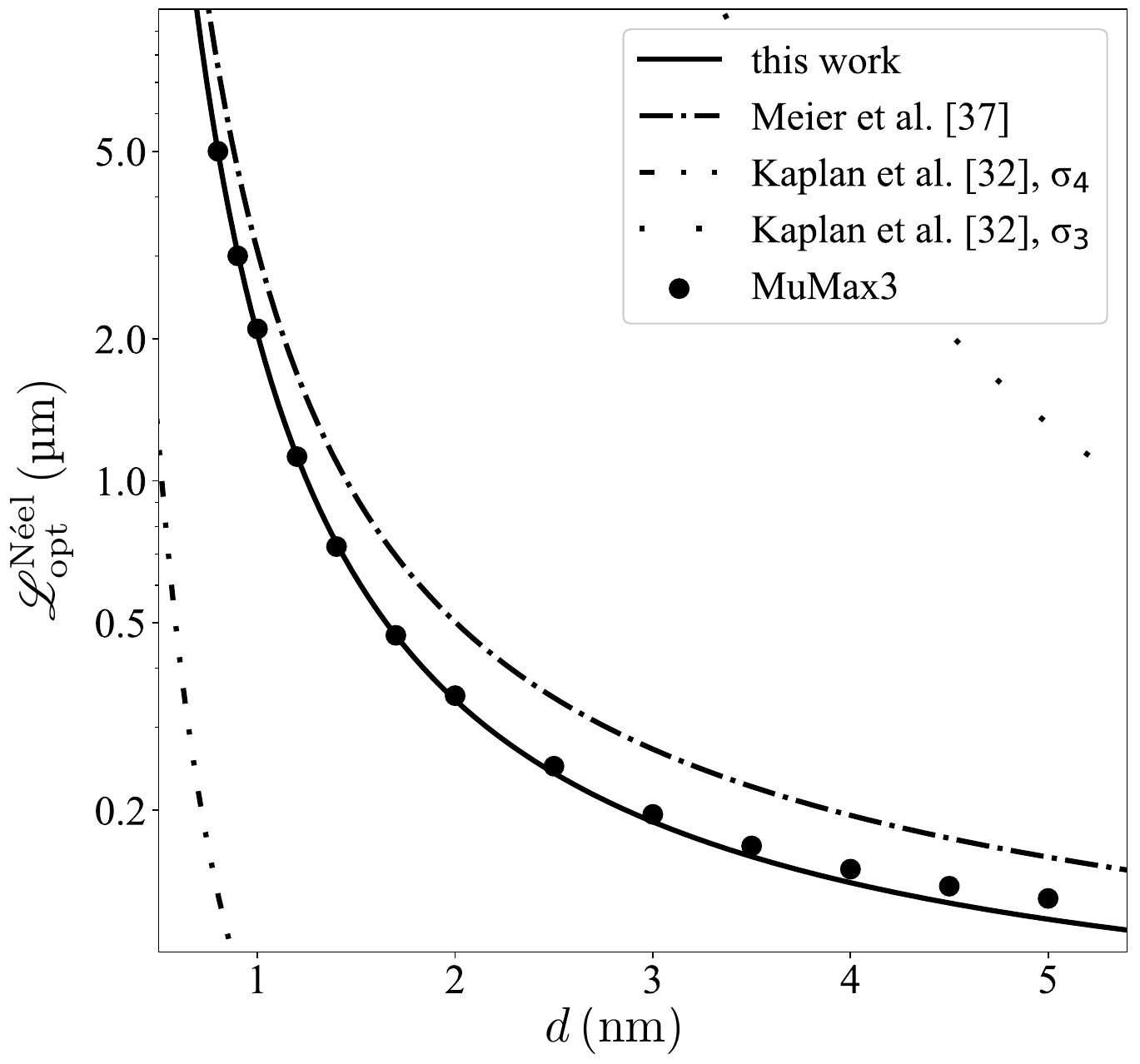}
\caption{Equilibrium stripe period
  ${\mathscr L}_\mathrm{opt}^\mathrm{N\acute{e}el}$ as a function of
  the film thickness $d$  in the regime of Néel walls. The thin film
  parameters are $A_\mathrm{ex}=10$ pJ$/$m, $M_s=1$ MA$/$m, $K_{u2}=1$
  MJ$/$m$^3$ and $D_2=2$ mJ$/$m$^2$ (regime 2). The asymptotic
  equilibrium stripe period in Eq. \eqref{LoptN} is represented by a
  solid line. The equilibrium stripe period obtained by micromagnetic
  simulations using {\sc MuMax3} (see Sec.~\ref{sec:micromag} for
  details) and in previous works\cite{meier17, kaplan93} are also
  presented (see inset for details).}
\label{fig:Neel}
\end{figure}

\begin{figure}[t]
\includegraphics[width=8.75cm]{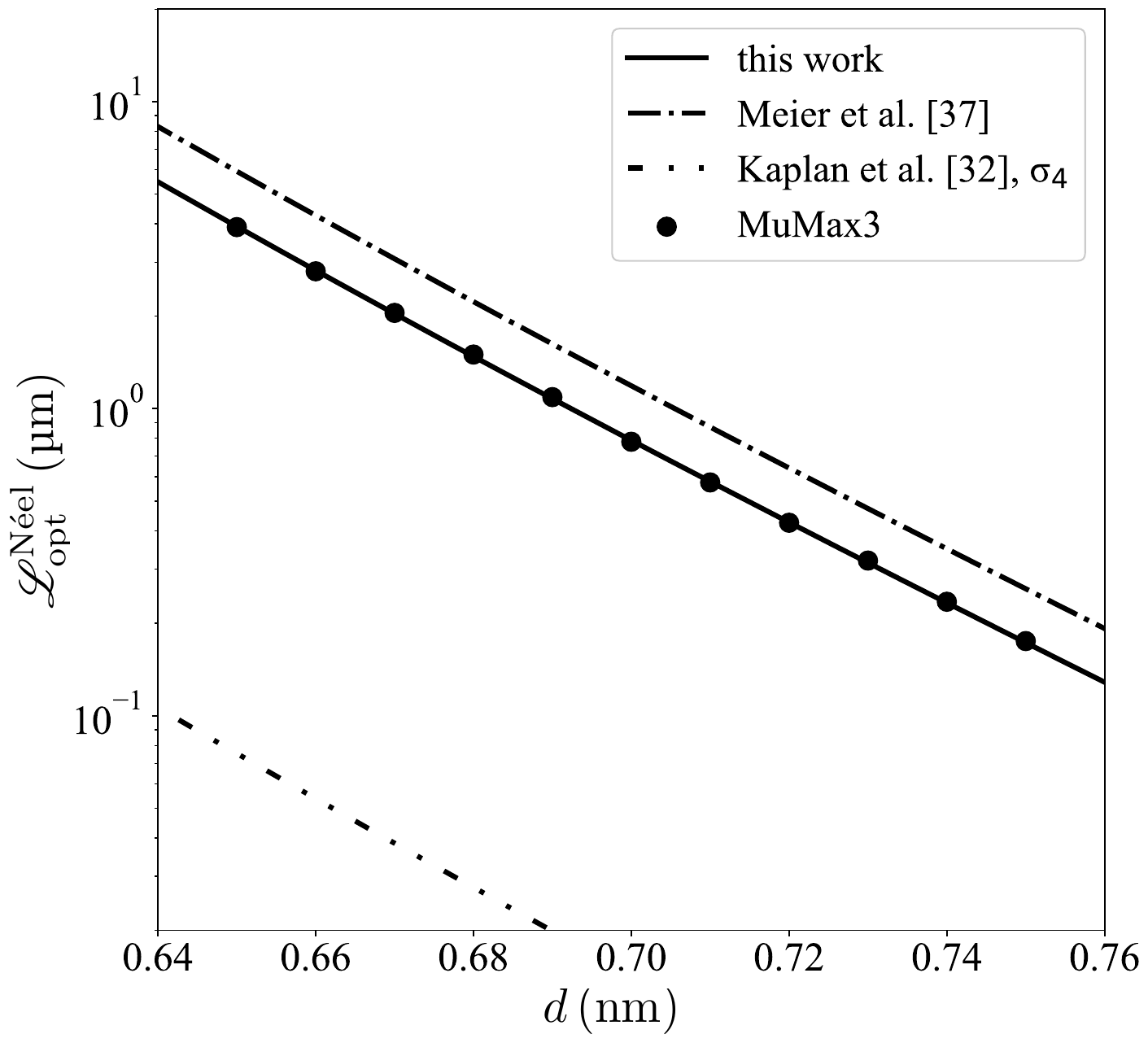}
\caption{Equilibrium stripe period
  ${\mathscr L}_\mathrm{opt}^\mathrm{N\acute{e}el}$ as a function of
  the film thickness $d$ in the regime of of Néel walls for an ultrathin
  film with thickness-dependent volume magnetocrystalline anisotropy
  $K_{u3}=K_{s3} / d$ and thickness-dependent volume DMI
  $D_{3}=D_{s3} / d$. The thin film parameters are $A_\mathrm{ex}=10$
  pJ$/$m, $M_s=1$ MA$/$m, $K_{s3}=0.6$ mJ$/$m$^2$, and $D_{s3}=1.2$
  pJ/m (regime 3). The asymptotic equilibrium stripe period in
  Eq. \eqref{LoptN} is represented by a solid line. The equilibrium
  stripe period obtained by micromagnetic simulations using {\sc
    MuMax3}\cite{vansteenkiste14} (see Sec.~\ref{sec:micromag} for
  details) and in previous works\cite{meier17, kaplan93} are also
  presented (see inset for details).}
\label{fig:atomic}
\end{figure}

The equilibrium stripe period formula for the case of Néel walls in
Eq. \eqref{LoptN} is illustrated by the solid line in
Fig.~\ref{fig:Neel}. We present the case of a magnetic thin film with
a volume magnetocrystalline anisotropy $K_{u2}=1$ MJ/m$^3$ and a
volume DMI $D_2=2$ mJ$/$m$^2$ (regime 2). Our asymptotic formula shows
a very good agreement with the micromagnetic simulations for low $d$
and starts to deviate as the thickness becomes greater than the Bloch
wall width $d>L_B\simeq 5$ nm, which is outside the range of validity
of our expression for the stray field energies valid in the ultrathin
film limit \cite{ms:prsa17,kmn:arma19}.  Finally, we address the
classical regime of a few monolayer transition metal wedge with
magnetocrystalline anisotropy and DMI of interfacial origin that are
inversely proportional to the film thickness: $K_{u3}=K_{s3} / d$ and
$D_{3}=D_{s3} / d$ (regime 3)\cite{heinrich93}. In this type of
systems, the observable stripe domains are present in a very narrow
range of thicknesses\cite{allenspach92,vaterlaus00,wu04,schott17} (a
fraction of a monolayer), as can be seen in Fig.~\ref{fig:atomic}. As
expected, the asymptotic model is very accurate in this regime, in
which $\delta \sim 0.1$. The variation of the equilibrium Néel stripe
period as a function of the DMI for a fixed thickness can be found in
the supplemental material \cite{suppl}.

\subsection{Comparison with previous studies}

In Figs.~\ref{fig:Bloch}--\ref{fig:atomic}, we also present the
equilibrium stripe periods predicted by existing formulas in the
literature and compare them with our asymptotic predictions in
Eqs.~\eqref{LoptB} and \eqref{LoptN}.  We start with the
Kaplan and Gehring formula that reads\cite{kaplan93,millev96}
\begin{align}
  \label{KG}
  \mathscr L_\mathrm{opt}^\mathrm{KG} = {\pi \over \sqrt{e}} \, d \exp
  \left( {\pi \sigma \over \mu_0 M_s^2 d} \right),
\end{align}
where $\sigma$ is the domain wall surface tension in three
dimensions. This formula has been widely used to obtain quantitative
information based on experimental measurement of the stripe period in
ultrathin films
\cite{hoffmann02,bergeard10,schott17,bouard18,schott21,balan23,je18}.
However, this formula is valid for films of thicknesses much larger
than the Bloch wall width, $d \gg L_B$, which is very unlikely to be
the case for an ultrathin film. Notice that if one were to use this
formula in the case of the Bloch stripes (no DMI), one would have to
choose $\sigma = \sigma_1 = 4 \sqrt{A_\mathrm{ex} K_u}$, as the
corresponding expression for the energy density already takes into
account the exact full magnetostatic energy of the three-dimensional
magnetic domains with zero wall thickness. Nevertheless, in the
literature one often assumed
$\sigma = \sigma_2 = 4 \sqrt{A_\mathrm{ex} (K_u - K_d)}$, which is not
consistent with the three-dimensional stray field energy calculation
used. For N\'eel stripes (with DMI), the correct choice of the domain
wall surface tension would be
$\sigma = \sigma_3 = 4 \sqrt{A_\mathrm{ex} (K_u + K_d)} - \pi D$ in
order to account for the additional stray field energy contribution
localized in the one-dimensional wall
profile\cite{bffflm:qam23}. Nevertheless,
$\sigma = \sigma_4 = 4 \sqrt{A_\mathrm{ex} (K_u - K_d)} - \pi D$ has
been frequently used. Regardless of the choice of $\sigma$, the
formula also suffers from the prefactor proportional to the film
thickness $d$.  In all the cases, the Kaplan and Gehring formula
presents, from one order of magnitude (regime 1, see
Fig.~\ref{fig:Bloch}), up to two orders of magnitude (regimes 2 and 3,
see Figs.~\ref{fig:Neel} and~\ref{fig:atomic}) errors in the predicted
equilibrium stripe period compared to our asymptotic formula, and its
use should, therefore, be avoided in the case of ultrathin films with
$d\lesssim L_B$.

The formulas of Kashuba and Pokrovsky \cite{kashuba93}, Sukstanskii
and Primak \cite{sukstanskii97}, and Skomski et. al \cite{skomski98}
are, on the contrary, valid in the ultrathin film limit, i.e., for
$d \lesssim L_B$. They are represented for the Bloch case in
Fig.~\ref{fig:Bloch}. They differ from Eq. \eqref{LoptB} only by the
numerical constant in front of the Bloch wall width in the prefactor,
since their prefactor is proportional to the Bloch wall width as in
Eq. \eqref{LoptB}. This constant is resulting from the long-range
dipolar interaction energy term for the stripe system, which is
calculated by evaluating a series (see
Sec.~\ref{sec:calculation}). Different assumptions or incorrect
evaluation of these series led to different prefactor constants in
those previous studies and to under- or overestimation of the
equilibrium stripe periods of less than an order of magnitude (see
Fig.~\ref{fig:Bloch}).

Finally, we compare our result with the work of Meier et
al. \cite{meier17}. In their work, Meier et al.  used the expression
given by Kashuba and Pokrovsky \cite{kashuba93} for the long-range
dipolar interaction energy of the stripe domains. They introduced
three modifications as compared to Kashuba and Pokrovsky.  First, the
expression of the domain wall surface tension appearing in the
exponential factor is modified to account for the wall surface tension
decrease by $\pi D$ in presence of DMI\cite{thiaville12}, in agreement
with Eq. \eqref{LoptN}. Second, they account for the increase in the
wall surface tension as the result of the presence of volume charges
in the case of a Néel wall\cite{tarasenko98}. This leads to an
additional factor of two in their prefactor as compared to the Bloch
case in agreement with the additional factor of 2 in Eq.~\eqref{LoptN}
as compared to Eq.~\eqref{LoptB}. Third, they introduce a
thickness-dependent domain wall width and domain wall surface
tension. One can see in Fig. \ref{fig:Bloch} that this last
modification introduced by Meier et al. leads to a slight decrease of
the predicted stripe period compared to that of Kashuba and Pokrovsky
in the case of the Bloch stripes and a slight apparent improvement of
the agreement with the numerics. Nevertheless, this improvement is
somewhat fortuitous, since the asymptotic behavior of the formula of
Meier et al. is the same as that of Kashuba and Pokrovsky, which
presents around a 50\% discrepancy with our exact asymptotic formula
and the results of direct numerical minimization of the micromagnetic
energy (see also Figs. \ref{fig:Neel} and \ref{fig:atomic} for the
case of the N\'eel stripes).  Our asymptotic formula, on the other
hand, yields a very good accuracy throughout the entire thickness
range as long as the film thickness is lower than the Bloch width
without the need to introduce any further thickness dependence related
to wall surface tension or domain wall width thickness variation (see
Figs. \ref{fig:Bloch}--\ref{fig:atomic}, see also
sec. \ref{sec:micromag} for a discussion on the domain wall width
thickness variation).

\section{Model}
\label{sec:Model}

We now present a detailed derivation of our formulas. We begin by
specifying the microscopic sample geometry. For simplicity, we
consider an extended material in the film plane and impose periodic
boundary conditions to avoid the need to deal with the material edge
effects (the treatment of the latter in the case of ultra-thin films
can be found in \cite{dms:mmmas24}).  Let
$\mathbb T_{{\mathscr L}}^2 = [0, {{\mathscr L}})^2$ be a flat
two-dimensional torus (a square box of sidelength ${\mathscr L}$ with
periodic boundary conditions) with ${\mathscr L}$ macroscopically
large. For a single layer of total thickness $d$ we define
$\widetilde \Omega = \mathbb T_\mathscr{L}^2 \times \left( -d/2, d/2
\right) \subset \mathbb R^3$ to be the set occupied by the
ferromagnetic layer.  Next we write the full micromagnetic energy
functional (in the SI units) for three-dimensional configurations
$\mathbf M = \mathbf M(x, y, z)$ described by the magnetization vector
$\mathbf M: \mathbb T_{{\mathscr L}}^2 \times \mathbb R \to \mathbb
R^3$ satisfying $|\mathbf M| = M_s$, the saturation magnetization (in
A/m), in $\widetilde \Omega$, and $\mathbf M = 0$ outside
$\widetilde \Omega$, with no applied field \cite{landau8,thiaville12}:
\begin{align}
  \label{EN}
  \begin{split}
    & \mathcal E^{3d}(\mathbf M) = \int_{\widetilde \Omega} \left(
      {A_\mathrm{ex} \over M_s^2} |\nabla \mathbf M|^2 + {K_u \over
        M_s^2} \left| \mathbf M_\perp \right|^2 \right)d^3 r
    \\
    & \qquad + {D \over M_s^2} \int_{\widetilde \Omega} \left( M_\|
      \nabla_\perp \cdot {\bf M}_\perp - {\bf M}_\perp \cdot
      \nabla_\perp M_\| \right) d^3 r \\
    & \qquad - {\mu_0\over2} \int_{\widetilde \Omega} \mathbf H_d
    \cdot \mathbf M \, d^3 r - \frac12 \mu_0 M_s^2 \mathscr L^2 d,
  \end{split}
\end{align}
where we used the notation
$\mathbf M = (\mathbf M_\perp, M_\parallel) \in \mathbb R^3$ with
$\mathbf M_\perp \in \mathbb R^2$ denoting the in-plane component of
$\mathbf M$ and $M_\parallel \in \mathbb R$ denoting the out-of-plane
component of $\mathbf M$ (parallel to the out-of-plane easy axis); the
operator $\nabla_\perp = (\partial_x, \partial_y)$ denotes the
in-plane portion of the gradient. The terms in the energy are, in
order of appearance: the exchange and the crystalline anisotropy with
exchange stiffness $A_\mathrm{ex}$ (in J/m, positive) and uniaxial
perpendicular anisotropy constant $K_u$ (in J/m$^3$, positive),
respectively; the interfacial DMI constant $D$ (in J/m$^2$, of
arbitrary sign); the stray field energy with the vacuum permeability
$\mu_0$ and $\mathbf H_d$ being the demagnetizing field solving the
static Maxwell's equations distributionally in
$\mathbb T_{\mathscr L}^2 \times \mathbb R$:
\begin{align}
  \label{Maxwell}
  \nabla \cdot (\mathbf H_d + \mathbf M) = 0, \qquad \nabla \times
  \mathbf H_d = 0,
\end{align}
with periodic boundary conditions in the plane and vanishing as
$z \to \pm \infty$; and an additive constant chosen so that
$\mathcal E^{3d}(\mathbf M) = 0$ for
$\mathbf M = \pm M_s \chi_{\widetilde \Omega} \hat{\mathbf z}$, where
$\chi_{\widetilde \Omega}$ is the characteristic function of
$\widetilde \Omega$, i.e., to offset the energy of the monodomain
state.

We next carry out a suitable non-dimensionalization. We measure all
lengths in the units of the exchange length
$\ell_\mathrm{ex} = \sqrt{2 A_\mathrm{ex} / (\mu_0 M_s^2)}$ and define
the normalized magnetization and energy
\begin{align}
  \label{mE}
  \mathbf m = {{\mathbf M} \over  M_s}, \qquad E^{3d} =
  {\mathcal E^{3d} \over  A_\mathrm{ex} d}.
\end{align}
Then the rescaled micromagnetic energy $E^{3d}$ is a function of
$\m : \mathbb T_L^2 \times \R \to \mathbb R^3$ with $|\m| = 1$ in
$\Omega = \mathbb T_L^2 \times (-\delta/2, \delta/2)$ and $|\m| = 0$
in the complement of $\Omega$, where
\begin{align}
  \label{Ldelta}
  L = {{\mathscr L} \over  \ell_\mathrm{ex}}, \qquad \delta =
  {d \over  \ell_\mathrm{ex}},  
\end{align}
and may be written as
\begin{align}
  \label{E}
  E^{3d}(\mathbf m)
  & = {1 \over \delta} \int_\Omega \left(
    |\nabla \m|^2 + Q |\mpe|^2 - 1 \right) d^3 r
    \notag \\ 
  & + {\kappa \over \delta} \int_\Omega \left( \mpa \nabla_\perp
    \cdot \mpe - 
    \mpe \cdot \nabla_\perp \mpa \right) \, d^3 r \\
  & + {1 \over \delta} \int_{\mathbb T_L^2 \times \mathbb R}  \nabla
    \cdot \m (-\Delta)^{-1} \nabla \cdot \m \, d^3 r. \notag
\end{align}
Here we introduced the dimensionless effective material quality factor
$Q > 1$ and the dimensionless DMI strength $\kappa$:
\begin{align}
  \label{Qkappah}
  \qquad Q = {2  K_u  \over \mu_0
  M_s^2 }, \qquad \kappa = D\sqrt{2  \over \mu_0
  M_s^2  A_\mathrm{ex}},
\end{align}
as well as the inverse Laplacian operator $(-\Delta)^{-1}$ whose
action on three-dimensional plane waves is defined as
\begin{align}
  \label{ivLapl}
  e^{-i \mathbf k \cdot \mathbf r}  (-\Delta)^{-1} e^{i \mathbf k
  \cdot \mathbf r} = {1 \over |\mathbf k|^2},
\end{align}
expressing the fact that the stray field energy represents the
Coulombic repulsive energy of the ``magnetic charges'', whose density
$\rho = -\nabla \cdot \m$ is understood in the distributional sense in
$\mathbb T_L^2 \times \mathbb R$ (Ref.\cite{dmrs:sima20}).  In the
rest of the paper we always consider the non-dimensional version of
the problem. For definiteness we also assume from now on that
$\kappa \geq 0$, without loss of generality.

We now introduce a reduced two-dimensional micromagnetic model
appropriate for ultra-thin films corresponding to $\delta \ll 1$.  We
define the magnetization $\m = \m(x, y)$ on a flat torus
$\mathbb T_L^2 = [0, L)^2$. The non-dimensional energy of such a
configuration in $\Omega$ is obtained via a suitable asymptotic
reduction from Eq.~\eqref{E} in the ultra-thin film limit (for
$\delta \ll 1$) and, up to an additive constant, reads
\cite{ms:prsa17,kmn:arma19}
\begin{align}
  \label{E2d}
  \begin{split}
    E_L^{2d}(\m) & = \int_{\mathbb T_L^2} \left( |\nabla \m|^2 + (Q -
      1) |\mpe|^2 \right) d^2 r \\ & + \kappa \int_{\mathbb T_L^2}
    \left( \mpa \nabla \cdot \mpe -
      \mpe \cdot \nabla \mpa \right) \, d^2 r  \\
    & - {\delta \over 2} \int_{\mathbb T_L^2} 
    \mpa (-\Delta)^{1/2} \mpa \, d^2 r \\ & + {\delta \over 2}
    \int_{\mathbb T_L^2} \nabla \cdot \mpe (-\Delta)^{-1/2} \, 
      \nabla \cdot \mpe \, d^2 r.
  \end{split}
\end{align}
Here the symbols $(-\Delta)^{1/2}$ and $(-\Delta)^{-1/2}$ denote the
half-Laplacian operator and its inverse, respectively, whose actions on
two-dimensional plane waves are defined as
\begin{align}
  \label{halfL2d}
  e^{-i \mathbf k \cdot \mathbf r} (-\Delta)^{1/2} e^{i \mathbf k
  \cdot \mathbf r}
  & = |\mathbf k|, \\
  e^{-i \mathbf k \cdot \mathbf r} (-\Delta)^{-1/2} e^{i \mathbf k
  \cdot \mathbf r}
  & = {1 \over |\mathbf k|}.
\end{align}
The terms associated with the half-Laplacian describe the nonlocal
contributions of the stray field due to the surface and volume
charges, respectively, with the usual local surface charge
contribution renormalizing the out-of-plane anisotropy constant. In
particular, we have $E^{2d}_L(\m) \simeq E^{3d}(\m \chi_\Omega)$ for
$\delta \ll 1$, where $\chi_\Omega$ is the characteristic function of
the set $\Omega$ and $\m \chi_\Omega$ represents a three-dimensional
magnetization configuration in the film that does not vary along the
$z$-direction.

\subsection{N\'eel stripes}
\label{sec:neel_model}

It may be conjectured, although it is a very difficult and unsolved
mathematical problem (for some recent developments, see
\cite{giuliani11,giuliani14,giuliani16,daneri19,daneri24}), that the
global energy minimizer of this energy with $\kappa$ and $\delta$
small, and with $L$ sufficiently large, is given by a periodic array
of stripes in which the magnetization rotates between the two easy
directions $\pm \hat{\mathbf z}$. In the presence of interfacial DMI
the rotation of the magnetization vector in the domain walls changes
its character from Bloch\cite{landau35,pratzer01} to
N\'eel\cite{heide06,heide08, tetienne15} as soon as
$\kappa \gtrsim \delta$, see\cite{thiaville12}. Hence, for
$\delta \ll 1$ and $\delta \lesssim \kappa \lesssim 1$ we may assume
that the magnetization profile is a one-dimensional N\'eel rotation
with angle $\theta = \theta(x)$, i.e., that
\begin{align}
  \label{mtheta}
  \m = (\sin \theta(x), 0, \cos \theta(x)).   
\end{align}
The energy of such a configuration is
$E_L^{2d}(\m) = L E_L^{1d}(\theta)$, where
\begin{align}
  \label{E1d}
  E_L^{1d}(\theta)
  & = \int_0^L \left( |\theta'|^2 + (Q - 1) \sin^2 \theta
    + \kappa \theta' \right) dx  \notag \\
  & - {\delta \over 2} \int_0^L 
    \cos \theta \left( - {d^2 \over dx^2} \right)^{1/2} \cos \theta \,
    dx \\
  & +{\delta \over 2} \int_0^L 
    \sin \theta \left( - {d^2 \over dx^2} \right)^{1/2} \sin \theta \,
    dx, \notag
\end{align}
where the fractional operator $\left( -{d^2 \over dx^2} \right)^{1/2}$
is the one-dimensional version of half-Laplacian, whose action on
one-dimensional plane waves is defined as
\begin{align}
  \label{halfact}
  e^{-i k x}  \left( -{d^2 \over dx^2} \right)^{1/2} e^{i k x} = |k|.
\end{align}
Note that this operator admits the following integral representation
\cite{mo:jcp06,mo:jap08}:
\begin{align}
  \label{halflapl}
  \left( -{d^2 \over dx^2} \right)^{1/2} u(x) = {1 \over \pi} \,
  \text{p.v.} \int_{-\infty}^\infty {u(x) - u(x') \over |x - x'|^2} \,
  dx' ,
\end{align}
where p.v. denotes the principal value of the integral at $x' = x$,
for any smooth periodic function $u$ extended to the whole real line.

The optimal configuration is obtained by minimizing the energy
$E_L^{2d}$ per unit area, hence if $\theta_L$ is a minimizer of
$E^{1d}_L$ and the stripe conjecture is valid, we have
\begin{align}
  \label{fL}
   f(L) = {1 \over L} E_L^{1d}(\theta_L) = \min_{\m \in H^1(\mathbb
  T_L^2; \mathbb S^2)} {1 \over L^2} E_L^{2d}(\m),
\end{align}
where the latter minimization is carried out over the standard class
$H^1(\mathbb T_L^2; \mathbb S^2)$ of three-dimensional vector fields
with values on a unit sphere and with square integrable weak
derivatives. A posteriori, any minimizer is known to be smooth and to
satisfy the Euler-Lagrange equation
\begin{align}
  \label{EL}
  \begin{split}
    0 = \theta_L''(x) - &(Q - 1) \sin \theta_L(x) \cos \theta_L(x) \\
    & - h_s(x) \sin \theta_L(x) + h_v(x) \cos \theta_L(x),
  \end{split}
\end{align}
where the effective fields $h_s$ and $h_v$ due to the long-range
dipolar contributions of the surface and volume charges are,
respectively,
\begin{align}
  \label{hshvhalf}
  h_s & = {\delta \over 2} \left( -{d^2 \over dx^2} \right)^{1/2} \cos
        \theta_L, \\
  h_v & = -{\delta \over 2} \left( -{d^2 \over dx^2} \right)^{1/2} \sin
        \theta_L, 
\end{align}
and the energy density $f(L)$ may be alternatively written as
\begin{align}
  \label{fL2}
  \begin{split}
   f(L) 
   = -{2 \pi \kappa \over L} + {1 \over L} \int_0^L &(
    |\theta_L'|^2 + (Q - 1) \sin^2 \theta_L\\
    &- h_v \sin \theta_L -  h_s
    \cos \theta_L ) dx .
    \end{split}
\end{align}

As $\cos \theta_L$ and $\sin \theta_L$ are assumed to be periodic,
without loss of generality and with a slight abuse of notation one can
take $L$ to be the fundamental period of the stripes, which we do from
now on. The DMI term forces winding of the angle by $-2 \pi$ over the
fundamental period for $\kappa > 0$, hence after a suitable
translation we may assume that $\theta_L(x)$ decreases from
$\theta_L(0) = {\pi \over 2}$ to $\theta_L(L) = -{3 \pi \over
  2}$. Thus Eq.~\eqref{EL} should be solved on the interval $(0,L)$
with the above Dirichlet boundary conditions at $x = 0$ and $x =
L$. Furthermore, as there is no applied magnetic field, we may assume
an additional symmetry $\theta_L({L \over 2} + x) = \theta_L(x) - \pi$
for $x \in (0, {L \over 2})$, consistent with Eq.~\eqref{EL}, which
allows to restrict the solution of Eq.~\eqref{EL} to the interval
$(0, {L \over 2})$ with Dirichlet boundary conditions
$\theta_L(0) = {\pi \over 2}$ and
$\theta_L\left( {L \over 2} \right) = -{\pi \over 2}$. In fact, one
further expects that $\theta_L$ is an odd function around
$x = {L \over 4}$ on $\left( 0, {L \over 2} \right)$, i.e.,
$\theta_L(x) = -\theta_L\left( {L \over 2}-x \right)$. In particular,
$\theta_L(x)$ decreases from $\theta_L(0) = {\pi \over 2}$ to
$\theta_L\left( {L \over 4} \right) = 0$ on
$\left( 0, {L \over 4} \right)$. Notice that these assumptions are
consistent with Eq.~\eqref{EL}, since $\cos \theta_L$ and
$\sin \theta_L$ being an even and odd function on
$\left( 0, {L \over 2} \right)$ around the midpoint $x = {L \over 4}$
results in $h_s$ and $h_v$ also being an even and odd functions,
respectively, consistent with $\theta_L''$ being an odd function.

\subsection{Bloch stripes}
\label{sec:bloch_model}

Here we briefly adapt the calculation above to the case of
$\kappa = 0$, no DMI. In this case the optimal stripe profile is of
Bloch type:
\begin{align}
  \label{mthetaB}
  \m = (0, \sin \theta(x), \cos \theta(x)),   
\end{align}
in order to minimize the contribution of bulk charges to the stray
field energy. As there is no contribution from bulk charges in this
case, the energy  is
\begin{align}
\begin{split}
  \label{E1dB}
  E_L^{1d}(\theta)
  = &\int_0^L  ( |\theta'|^2 + (Q - 1) \sin^2 \theta  ) dx \\
  & -   {\delta \over 2} \int_0^L  
  \cos \theta \left( - {d^2 \over dx^2} \right)^{1/2} \cos \theta \,
  dx,
  \end{split}
\end{align}
and the Euler-Lagrange equation satisfied by an associated minimizer
$\theta_L$ is
\begin{align}
  \label{ELB}
  0 = \theta_L''(x) - (Q - 1) \sin \theta_L(x) \cos
  \theta_L(x) \notag \\ 
  -  h_s(x) \sin \theta_L(x). 
\end{align}

Contrary to the case of strong DMI, with only the surface charge
contribution to the energy present the optimal profile $\theta_L(x)$
is expected to exhibit no winding and hence be periodic (see also
\cite{giuliani09}). The energy density of stripes is then given by
\begin{align}
  \label{fL2B}
   f(L) 
  = {1 \over L} \int_0^L \left(
  |\theta_L'|^2 + (Q - 1) \sin^2 \theta_L -  h_s
  \cos \theta_L 
  \right) dx.
\end{align}


\section{Calculation of the optimal stripe period in ultrathin films}
\label{sec:calculation}

We now investigate the minimization problem associated with $ f(L)$ in
the limit $\delta \to 0$. It is known rigorously \cite{kmn:arma19}
that in this regime (with the effect of not too strong DMI readily
incorporated as in \cite{ms:prsa17}) the average length scale of the
energy-minimizing patterns for Eq.~\eqref{E2d} on all sufficiently
large spatial domains scales as
\begin{align}
  \label{Lscale}
  L \sim {e^{a / \delta} \over \sqrt{Q - 1}}, \qquad a = \frac{\pi}{2} \left( 4
  \sqrt{Q - 1} - \pi \kappa \right) > 0.
\end{align}
In particular, under the periodic stripe conjecture Eq.~\eqref{Lscale}
gives the leading order scaling (up to a prefactor) of the optimal
stripe period. Furthermore, if $L$ is the fundamental period, we have
$\left| E_L^{1d}(\theta_L) \right| \leq C$ for some constant $C > 0$
independent of $\delta$, and the same estimate holds for every term in
the energy separately \cite{kmn:arma19}.

\subsection{N\'eel stripes period}
\label{sec:neel_calc}

To calculate the optimal period, we first need to approximate the
energy density $ f(L)$ for sufficiently small values of $\delta$. This
requires to find a leading order approximation to the minimizing
profile $\theta_L$, which in view of the fact that $L \to \infty$ as
$\delta \to 0$ is, in principle, a singular perturbation
problem. Nevertheless, at least formally it is possible to show that
the leading order behavior of $\theta_L$ on the interval
$\left( 0, {L \over 4} \right)$ may be obtained by setting
$\delta = 0$ in Eq.~\eqref{EL}, yielding a unique monotonically
decreasing solution $\theta_L^{(0)}$ satisfying
$\theta_L^{(0)}(0) = {\pi \over 2}$ and
$\theta_L^{(0)} \left( {L \over 4} \right) = 0$.  Indeed, from the
integral representation of $\left( -{d^2 \over dx^2} \right)^{1/2}$ in
Eq.~\eqref{halflapl} we have for any function
$u \in C^\infty(\mathbb R) \cap L^\infty(\mathbb R)$
\begin{align}
  \label{halfbreak}
  \left( - {d^2  \over dx^2} \right)^{1/2} u(x)
    = {1 \over \pi} \,
    \text{p.v.} \int_{x-1}^{x+1} {u(x) - u(x') \over |x - x'|^2} \,
    dx' \notag \\
    + {1 \over \pi} \int_{-\infty}^{x-1} {u(x) - u(x') \over |x - x'|^2} \,
    dx' + {1 \over \pi} \int_{x+1}^\infty {u(x) - u(x') \over |x - x'|^2} \,
    dx'.
\end{align}
Therefore
\begin{align}
  \label{halfuinfty}
  \left\|  \left( - {d^2  \over dx^2} \right)^{1/2} u
  \right\|_{L^\infty(\mathbb R)} \leq {4 \over \pi} \| u
  \|_{L^\infty(\mathbb R)} + {1 \over \pi} \| u'' \|_{L^\infty(\mathbb
  R)}.  
\end{align}
Thus, with $u = \cos \theta_L$ or $u = \sin \theta_L$ we have
\begin{align}
\begin{split}
  \label{hshvinfty}
  \| h_s &\|_{L^\infty(\mathbb R)} + \| h_v \|_{L^\infty(\mathbb R)}\\
  &
  \leq C \delta \left( 1 + \| \theta_L' \|_{L^\infty(\mathbb R)}^2 + \|
  \theta_L'' \|_{L^\infty(\mathbb R)} \right),
  \end{split}
\end{align}
for some universal constant $C > 0$. Furthermore, arguing as in
\cite{my:prsa16}, one can show that
$\| \theta_L' \|_{L^\infty(\mathbb R)}$ and
$\| \theta_L'' \|_{L^\infty(\mathbb R)}$ are both uniformly bounded
for all bounded solutions of Eq.~\eqref{EL}, hence the stray field
terms are uniformly small, and one can therefore look for a solution
of the Dirichlet problem for Eq.~\eqref{EL} in the form of a regular
series expansion in $\delta$. Furthermore, it is not difficult to see
directly from Eq.~\eqref{EL} with $\delta = 0$ that
\begin{align}
  \label{thL0}
  \theta_L^{(0)}(x) = \arccos \left[ \tanh \left( x \sqrt{Q - 1}
  \right) \right] + O(e^{-cL}),
\end{align}
for some $c > 0$ when $L \gg 1$, for all
$x \in \left( 0, {L \over 4} \right)$.

Having formally established that the minimizer
$\theta_L \simeq \theta_L^{(0)}$ to within $O(\delta)$ accuracy, we
proceed to calculate to the leading order in $\delta$: 
\begin{align}
\begin{split}
  \label{fL0}
  & f(L) = -{2 \pi \kappa \over L} \\
  &+ {4 \over L} \int_0^{L/4} \left( \left| {d \theta_L^{(0)} \over
        dx} \right|^2 + (Q - 1) \sin^2
    \theta_L^{(0)} \right) dx  \\
  & + {2 \delta \over L} \int_0^{L/4} \left( \sin \theta_L^{(0)}
    \left( - {d^2 \over dx^2} \right)^{1/2} \sin \theta_L^{(0)}
  \right) dx \\
  &- {2 \delta \over L} \int_0^{L/4} \left(\cos \theta_L^{(0)} \left(
      - {d^2 \over dx^2}
    \right)^{1/2} \cos \theta_L^{(0)}\right) dx \\
  & + {O(\delta^2) \over L} ,
  \end{split}
\end{align}
where the $O(\delta^2)$ error term arises from the fact that
$\theta_L^{(0)}$ is a strict local minimizer of $E_L^{1d}$ with
$\delta = 0$ and the same boundary data, and that
$\theta_L - \theta_L^{(0)} = O(\delta)$. Furthermore, as the
perturbations are localized in the vicinity of the domain walls, this
error term is expected to be uniform in $L$. Indeed, with the
minimizer $\theta_L$ close to $\theta_L^{(0)}$, by
Eq.~\eqref{halflapl} we have
\begin{align}
  \label{hshvdecay}
  h_s(x) \sim {\delta \over x}, \qquad h_v(x) \sim {\delta \over x^2},
  \qquad 1 \ll x \ll L.
\end{align}
Hence the perturbation effectively vanishes in the space between the
domain walls.

We now write $ f(L) \simeq f_1(L) + f_2(L)$, where $f_1$ is the sum of
the first and the second lines, and $f_2$ is the sum of the third and
the fourth lines in the right-hand side of Eq.~\eqref{fL0}. An
explicit calculation shows that
\begin{align}
  \label{f1}
   f_1(L) = {2 \over L} \left( 4
  \sqrt{Q - 1} - \pi \kappa \right) + O\left( e^{-cL} \right),
\end{align}
for some $c > 0$ and all $\delta \ll 1$. This is just the energy of
two N\'eel walls per period, up to the exponential order in $L \gg 1$.

To calculate $f_2$, we pass to the Fourier series representations of
$\cos \theta_L^{(0)}$ and $\sin \theta_L^{(0)}$. From the symmetries
of $\theta_L^{(0)}$ we have
$\cos \theta_L^{(0)}(x) = \cos \theta_L^{(0)} \left( \frac{L}{2} - x
\right)$ and
$\sin \theta_L^{(0)}(x) = -\sin \theta_L^{(0)}\left( \frac{L}{2} - x
\right)$ for $x \in [\frac{L}{4}, \frac{L}{2}]$, and
$\cos \theta_L^{(0)}(x) = -\cos \theta_L^{(0)}(L - x)$ and
$\sin \theta_L^{(0)}(x) = \sin \theta_L^{(0)}(L - x)$ for
$x \in [\frac{L}{2}, L]$. Defining the Fourier series of the resulting
functions, we then obtain
\begin{align}
  \label{sincosF}
  \sin \theta_L^{(0)}(x)
  & = \sum_{m=1}^\infty a_{2m-1} \cos \left(
    { 2 \pi (2m-1) x \over L} \right), \\
  \cos \theta_L^{(0)}(x)
  & = \sum_{m=1}^\infty b_{2m-1} \sin \left(
    { 2 \pi (2m-1) x \over L} \right). \label{sincoscosF}
\end{align}

Using the fact that by Eq.~\eqref{thL0} we have
\begin{align}
  \label{cossin}
  \cos \theta_L^{(0)}(x)
  & \simeq \tanh \left( x \sqrt{Q - 1} \right), \qquad
    x \in \left[ 0, \tfrac{L}{4} \right], \\
  \sin \theta_L^{(0)} (x)
  & \simeq \mathrm{sech} \left( x \sqrt{Q - 1}
    \right), \qquad x \in \left[ 0, \tfrac{L}{4} \right],
\end{align}
up to $O(e^{-cL})$ errors, the coefficients of the sine series can be
computed as
\begin{align}
  \label{bn}
  b_{2m - 1}
  & =  {4 \over L} \int_0^{L/2} \cos \theta_L^{(0)}(x)
    \sin \left(
    { 2 \pi (2 m - 1) x \over L} \right) \, dx \notag \\
  & = - {2 \over (2 m - 1) \pi} \notag \\
  \times \int_0^{L/2} & {d \theta_L^{(0)} (x) \over dx} \sin
                        \theta_L^{(0)}(x)  \cos \left(
                        { 2 \pi (2 m - 1) x \over L} \right) \, dx \notag \\
  & = - {4 \over (2 m - 1) \pi} \notag \\
  \times \int_0^{L/4} & 
                        {d \theta_L^{(0)} (x) \over dx} \sin
                        \theta_L^{(0)}(x)  \cos \left(
                        { 2 \pi (2 m - 1) x \over L} \right) \, dx  \\
  & \simeq {4 \sqrt{Q - 1} \over (2 m - 1) \pi} \notag \\
  \times \int_0^\infty & 
                         \mathrm{sech}^2 \left( x \sqrt{Q - 1} \right)  \cos \left(
                         { 2 \pi (2 m - 1) x \over L} \right) \, dx
                         \notag \\
  & =  {4 \pi \over L \sqrt{Q - 1}} \, 
    \mathrm{csch} \left( { (2 m - 1) \pi^2 \over L \sqrt{Q - 1} }
    \right), \notag 
\end{align}
where we integrated by parts in the first line and used the symmetry
of $\cos \theta_L^{(0)}$ in the second line, again with errors of
exponential order. Similarly, the coefficients of the cosine series
are
\begin{align}
  \label{an}
  a_{2m-1}
  &  =  {8 \over L} \int_0^{L/4} \sin \theta_L^{(0)}(x)
    \cos \left(
    { 2 \pi (2 m - 1) x \over L} \right) \, dx \notag \\
  & \simeq {8 \over L} \int_0^\infty \mathrm{sech} \left( x
    \sqrt{Q - 1} \right) 
    \cos \left(
    { 2 \pi (2 m - 1) x \over L} \right) \, dx \notag \\
  & = {4 \pi \over L \sqrt{Q - 1}} \,  \mathrm{sech} \left( { (2 m -
    1) \pi^2 \over L \sqrt{Q - 1} } 
    \right). 
\end{align}

Finally, with the help of Eq.~\eqref{halfact} and Eq.~\eqref{fL0} the stray
field energy density is
\begin{align}
  \label{f2F}
  f_2(L)
  & \simeq - {\pi \delta \over 2 L} \sum_{m=1}^\infty (2 m -
    1) \left( |b_{2m-1}|^2 - |a_{2m-1}|^2 \right) \notag \\
  & = -{32 \pi^3 \delta \over L^3 (Q - 1)} \notag \\
  & \times \sum_{m=1}^\infty (2 m -
    1) \mathrm{csch}^2  \left( { 2 \pi^2 (2 m - 1) \over L \sqrt{Q - 1} }
    \right). 
\end{align}

Let us define $\iks = \frac{2\pi^2}{L \sqrt{Q - 1}}$ and compute the
above sum as
\begin{align}
  & \sum_{m=1}^\infty \frac{2m -1}{\sinh^2 [(2m-1)\iks]}  \notag \\
  & =
    \sum_{m=1}^\infty \frac{m}{\sinh^2( m \iks)} -  \sum_{m=1}^\infty 
    \frac{2m}{\sinh^2( 2m \iks)},
\end{align}
by splitting the series in the first term in the right-hand side into
the even and odd terms.  Since for $\iks>0$ the series above converge
exponentially fast, we have
\begin{align}
  & \sum_{m=1}^\infty \frac{2m -1}{\sinh^2[(2m-1)\iks]} =
    \frac{d}{d\iks} \sum_{m=1}^\infty (1-\coth(m\iks))  \notag \\
  & \qquad \qquad -
    \frac{d}{d\iks} \sum_{m=1}^\infty (1-\coth(2m\iks)). 
\end{align}
Hence it is enough to compute the series
\begin{align}
  & \sum_{m=1}^\infty  (1-\coth(m\iks)) \notag \\
  & \qquad = 2 \sum_{m=1}^\infty
    \frac{e^{-2m\iks}}{e^{-2m\iks}-1} = -2 \mathcal L (e^{-2\iks}), 
\end{align}
where $\mathcal L(q)= \sum_{m=1}^\infty \frac{q^m}{1-q^m }$ is a
well-studied Lambert series (see e.g. \cite{banerjee17}). We obtain
\begin{align}
  \label{eq:f1}
  & \sum_{m=1}^\infty \frac{2m -1}{\sinh^2[(2m-1)\iks]}  \notag \\
  & \qquad = -2
    \frac{d}{d\iks} 
    (\mathcal L (e^{-2\iks}) -\mathcal L (e^{-4\iks})). 
\end{align}
When $\iks \to 0$ and, hence, $q=e^{-2\iks} \to 1^-$, we have the
following asymptotic expansion of $\mathcal L(q)$ (see
e.g. \cite[Theorem 2.2(2)]{banerjee17}, and \cite{kluyver19}): 
\begin{align}
  \label{eq:f2}
  \mathcal L(q)
  & = \sum_{m=1}^\infty  \frac{q^m}{1-q^m} \notag \\
  & 
    = -\frac{\ln\ln\frac{1}{q} -\gamma}{\ln\frac{1}{q}} +\frac14 -
    \sum_{n=1}^\infty  \frac{B_n^2}{n\, n!} \left(\ln\frac{1}{q}
    \right)^n  ,
\end{align}
where $\gamma$ is the Euler-Mascheroni constant and $B_n$ is the
$n$-th Bernoulli number. Therefore, for $\iks \to 0$, using
Eqs.~\eqref{eq:f1} and \eqref{eq:f2}, we obtain
\begin{align}
  \sum_{m=1}^\infty \frac{2m -1}{\sinh^2[(2m-1)\iks]} = \frac{
  \gamma+1 -\ln
  \iks }{2\iks^2} +O(1). 
  \end{align}
  Recalling that $\iks =\frac{2\pi^2}{L\sqrt{Q - 1}}$, for $L \gg 1$
  and $\delta \ll 1$ we then obtain
  \begin{align}
    \label{f2est}
  f_2(L)
  =  -{4 \delta \over \pi L}  \bigg[ \ln
  & \left(\frac{L\sqrt{Q
    - 1}}{2\pi^2} \right) \notag \\
  & +\gamma+1 + O(1/L^2) + O(\delta)  \bigg]. 
\end{align}
for some $c > 0$.  Thus, within the
$O(\delta^2 / L) + O(\delta / L^3) + O(e^{-cL})$ errors the total
energy density is
\begin{align}
  \label{energy_neel}
  f(L) \simeq {2 \over L}
  & \left( 4
    \sqrt{Q - 1} - \pi \kappa \right) \notag \\
  & -{4 \delta \over \pi L}
    \left[\ln\left(\frac{L\sqrt{Q - 1}}{2\pi^2}\right) +\gamma+1
    \right]. 
\end{align}

We now minimize the obtained expression in $L$. A simple calculation
shows that the expression for $ f(L)$ in Eq.~\eqref{energy_neel} is
uniquely minimized by $L = L_\mathrm{opt}$, where
\begin{align}
  L_\mathrm{opt} \simeq {2 \pi^2 e^{-\gamma} \over \sqrt{Q - 1}} \exp 
  \left[ {\pi \over 2 \delta} \left( 4 \sqrt{Q - 1} - \pi \kappa
  \right) \right]. 
\end{align}
As expected, the optimal period is of the form appearing in
Eq.~\eqref{Lscale}, which justifies neglecting the error terms in the
expression for $ f(L)$ as $\delta \to 0$, and gives the leading order
asymptotic behavior of the optimal stripe period in this limit.

\subsection{Bloch stripes period}
\label{sec:bloch_calc}

We can follow the same arguments as in Sec. \ref{sec:neel_calc}
applied to the Bloch stripes for $\kappa = 0$ discussed in
Sec. \ref{sec:bloch_model}. As in the case of the N\'eel stripes,
  the leading order behavior of the minimizer $\theta_L$ on the
  interval $(0,{L \over 4})$ may be formally obtained by setting
  $\delta = 0$ in Eq.~\eqref{ELB}, yielding a unique solution
  $\theta_L^{(0)}$ satisfying $\theta_L^{(0)}(0) = {\pi \over 2}$ and
  $\theta_L^{(0)} \left( {L \over 4} \right) = 0$. This solution again
  satisfies the estimate in Eq.~\eqref{thL0}. Proceeding as in
  Eq.~\eqref{fL0}, we then obtain that
\begin{align}
\begin{split}
  \label{fL0B}
  & f(L) = {4 \over L} \int_0^{L/4} \left( \left| {d \theta_L^{(0)}
        \over dx} \right|^2 + (Q - 1) \sin^2
    \theta_L^{(0)} \right) dx  \\
  &- {2 \delta \over L} \int_0^{L/4} \left(\cos \theta_L^{(0)} \left(
      - {d^2 \over dx^2}
    \right)^{1/2} \cos \theta_L^{(0)}\right) dx \\
  & + {O(\delta^2) \over L},
  \end{split}
\end{align}
and we can write $f(L) \simeq f_1(L) + f_2(L)$, where
\begin{align}
  \label{f1B}
  f_1(L) = {8 \sqrt{Q - 1} \over L} + O\left( e^{-cL} \right),
\end{align}
and the stray field contribution $f_2$ is calculated by passing to the
Fourier series. Using the fact that $\cos \theta_L^{(0)}$ is an odd
$L$-periodic function, we again obtain the expression in
Eq.~\eqref{sincoscosF}, where the Fourier coefficients are given by
Eq.~\eqref{bn}.  Using Eqs.~\eqref{halfact} and \eqref{fL0B}, we then
arrive at the formula
\begin{align}
 \label{f2FB}
  f_2(L)
  & \simeq -{8 \pi^3 \delta \over L^3 (Q - 1)} \notag \\
  & \times \sum_{m=1}^\infty (2 m
    - 1) \mathrm{csch}^2 \left( { \pi^2 (2 m - 1) \over L \sqrt{Q - 1} }
    \right).
\end{align}

An expansion analogous to the one in Sec. \ref{sec:neel_calc} can be
obtained by noting that the expression in the right-hand side of
Eq.~\eqref{f2FB} is equal to twice the one in Eq.~\eqref{f2F} with $L$
replaced by $2L$. Hence from Eqs.~\eqref{f1B} and \eqref{f2est}
we can infer that
\begin{align}
 \label{energy_bloch}
  f(L) \simeq
  & {8 \sqrt{Q - 1} \over L}
    \notag \\
  & -{4 \delta \over \pi L}
    \left[\ln\left(\frac{L\sqrt{Q - 1}}{\pi^2}\right) +\gamma+1
    \right]. 
\end{align}
Optimizing this expression then gives, to the leading order, the
optimal period
\begin{align}
  L_\mathrm{opt} \simeq {\pi^2 e^{-\gamma} \over \sqrt{Q - 1}} \exp
  \left( {2 \pi \over \delta} \sqrt{Q - 1} \right). 
\end{align}
\section{Micromagnetic simulations}
\label{sec:micromag}
We carried out detailed micromagnetic simulations of the magnetic
stripe domains, using the open source {\sc MuMax3} software
\cite{vansteenkiste14}. Our system consists of a box of
$N_x\times 256 \times 1$ cells, where $N_x$ is varied. The individual
cell size is $l_x\times4$ nm $\times $ $l_z$, where $l_x$ is set to
$1$ nm in the Bloch case (regime 1) and $0.25$ nm in the Néel case
(regimes 2 and 3), and $l_z = d$, where $d$ is varied between $d=0.65$
nm ($\delta\simeq 0.16$) and $d=8$ nm ($\delta\simeq 2$). The choice
of the number of cells in the $y$-direction is dictated by the {\sc
  MuMax3} implementation of the periodic boundary conditions for the
stray field, which are approximated by specifying the number of
repeats of the computational domain to define the magnetostatic
kernel. We set the number of repeats in $(X,Y,Z)$ to (5,5,0) alongside
with our choice of the discretization cell sizes and numbers in order
to achieve a sufficiently good accuracy of the calculation.

We fix the system exchange constant to $A_\mathrm{ex}=10$ pJ/m and
saturation magnetization to $M_s=1$ MA/m, resulting in an exchange
length $\ell_\mathrm{ex} \simeq 4 $ nm. A snapshot of the simulated
system in the case of Néel walls (regime 2), which has been minimized
in energy, is shown in Fig.~\ref{fig:Schem}(a).  In this simulation,
the stripe period is imposed. As a consequence, in order to find the
lowest energy configuration, we repeat the simulation for various
$N_x$ and look for a minimum of the energy per unit volume as a
function of $N_x$. The equilibrium stripe period predicted by
Eqs. \eqref{LoptB} and \eqref{LoptN} serves as a guide to guess the
period in the simulation, and the simulations converge quickly to the
minimum of energy for each $N_x$.

\begin{figure}[t]
\includegraphics[width=8.5cm]{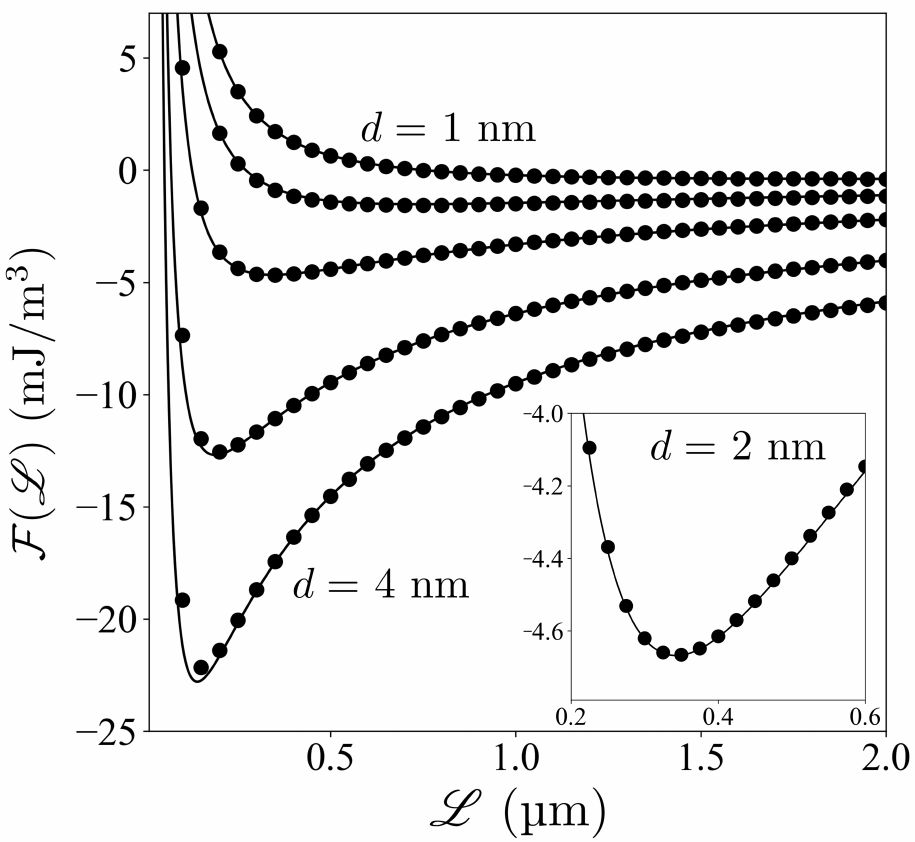}
\caption{Comparison between the numerical simulations in dots ({\sc
    MuMax3}\cite{vansteenkiste14}) and dimensional version of the
  asymptotic energy
  ${\mathcal F}({\mathscr L}) = f(\mathscr L / \ell_\mathrm{ex}) K_d$,
  where $f(L)$ can be found in Eq. \eqref{energy_neel}, represented as
  a solid line. The parameters are $A_\mathrm{ex}=10$ pJ/m, $M_s=1$
  MA/m, $K_{u2}=1$ MJ$/$m$^3$ and $D_2=2$ mJ$/$m$^2$. The film
  thicknesses are (starting from the bottom curve) $d=4$ nm, $d=3$ nm,
  $d=2$ nm, $d=1.4$ nm and $d=1$ nm.}
\label{fig:singleEL}
\end{figure}

We extract the stripe system energy density from the simulation by
suitably offsetting the total energy density output parameter of {\sc
  MuMax3} and compare it to the dimensional asymptotic energy density
${\mathcal F}({\mathscr L}) = f(\mathscr L / \ell_\mathrm{ex}) K_d$ in
mJ/m$^3$, where $f(L)$ is the energy density of the stripe system
computed in Sec.~\ref{sec:calculation}, see Eq. \eqref{energy_neel}
for the Néel case and Eq. \eqref{energy_bloch} for the Bloch case. In
Fig. \ref{fig:singleEL}, we show, for regime 2 (Néel rotation,
$K_{u2}=1$ MJ$/$m$^3$ and $D_2=2$ mJ$/$m$^2$), the energy density
(dots) obtained numerically using the {\tt minimize} routine for the
values of $N_x l_x$ between 100 nm and 2 µm and the thickness $d$
varying from 1 nm to 4 nm. The dimensional asymptotic energy density
${\mathcal F}({\mathscr L}) = f(\mathscr L / \ell_\mathrm{ex}) K_d$
obtained from Eq. \eqref{energy_neel} and represented by the solid
line in Fig.~\ref{fig:singleEL} shows excellent agreement with the
numerical simulations.  The good agreement between the asymptotics and
the simulations starts to deviate around $d\simeq 4$ nm
($d \simeq L_B$).

We want to mention that in a ferromagnetic thin film, the domain wall
width has been shown to be thickness-dependent, even in the ultrathin
film limit \cite{lemesh17} ($d \to 0$). In regime 1, the domain wall
width estimated from a fit to the profile obtained with the
micromagnetic simulations presents, for $d=8$ nm, a 40\% decrease
compared to $L_B$. However, the excellent agreement between the
micromagnetic simulations and the asymptotic formulas in this work
shows that the equilibrium stripe period is unaffected by this
dependence to the leading order.

\section{Conclusions}
\label{sec:concl}

We have derived analytical formulas for the equilibrium stripe period
which present an inverse exponential dependence on the film thickness
and a prefactor proportional to the Bloch wall width. These formulas
are derived for the case of pure Néel and Bloch rotations and are
asymptotically exact for vanishing film thicknesses in the presence or
absence of interfacial DMI, respectively. The formulas are applied to
classical sets of system parameters, including the Bloch and the Néel
regimes and a film thickness varying from one monolayer to about 10
nm.  The comparison with micromagnetic simulations in the considered
regimes shows excellent agreement, confirming the applicability of our
formula in the ultrathin film regime.  This agreement represents a
quantitative improvement as compared to the state of the art formulas
in the literature as previous studies did not succeed to obtain the
correct asymptotics for vanishing thicknesses.  The accuracy of our
formulas is remarkably robust up to thicknesses of at least twice the
exchange length. The variation of the wall width with the film
thickness does not affect this accuracy and the formulas remain
accurate as long as the thickness is smaller than the Bloch wall
width. We also highlight the inapplicability, in the ultrathin film
regime, of an alternative formula obtained by neglecting the wall
width (thin wall approximation) which leads to a prefactor
proportional to the film thickness. This settles a controversy with
regards to which formula should be used for predicting the equilibrium
stripe period in ultrathin films. We explicitly clarify the conditions
of validity of the respective formulas as well as the proper choice of
domain wall surface tension in the different regimes.

\begin{acknowledgments}
  A. Bernand-Mantel was supported by France 2030 government investment
  plan managed by the French National Research Agency under grant
  reference PEPR SPIN [SPINTHEORY] ANR-22-EXSP-0009 and grant NanoX
  ANR-17-EURE-0009 in the framework of the Programme des
  Investissements d'Avenir. C. B. Muratov was supported by MUR via
  PRIN 2022 PNRR project P2022WJW9H and acknowledges the MUR
  Excellence Department Project awarded to the Department of
  Mathematics, University of Pisa, CUP I57G22000700001. C. B. Muratov
  is a member of INdAM-GNAMPA. A. Bernand-Mantel thanks N. Reyren and
  W. Legrand for fruitful discussions.
\end{acknowledgments}

\end{document}